\documentclass[aip]{revtex4-1} 

\usepackage[utf8]{inputenc}    
\usepackage[T1]{fontenc}       

\usepackage{graphicx} 
\usepackage{epstopdf, epsfig}
\graphicspath{ {./Figures/}}
\usepackage{overpic}    
\AppendGraphicsExtensions{.tiff}

\usepackage{multirow} 
\usepackage{booktabs} 
\usepackage{array} 

\usepackage{natbib}            
\usepackage{hyperref}          
\usepackage[capitalise]{cleveref} 
\Crefname{figure}{Fig.}{Figs.} 

\usepackage{lineno} 
\usepackage{xcolor} 
\usepackage{soul} 

\begin{document}

\preprint{JRSE20-AR-00951}

\scriptsize
\noindent\textcolor{blue}{This article may be downloaded for personal use only. Any other use requires prior permission of the author and AIP Publishing. This article appeared in Hunt \textit{et al}, J. Renewable Sustainable Energy \textbf{12}, 054501 (2020) and may be found at \url{https://doi.org/10.1063/5.0016753}.}
\normalsize

\title[Effect of Aspect Ratio on Cross-Flow Turbine Performance]{Effect of Aspect Ratio on Cross-Flow Turbine Performance}

\author{Aidan Hunt}
    \email{ahunt94@uw.edu} 
    \affiliation{Department of Mechanical Engineering, University of Washington}
\author{Carl Stringer}
    \affiliation{Department of Mechanical Engineering, University of Washington}
\author{Brian Polagye}
    \affiliation{Department of Mechanical Engineering, University of Washington}
    
\date{6 August 2020}

\begin{abstract}
    \noindent Cross-flow turbines convert kinetic power in wind or water currents to mechanical power. Unlike axial-flow turbines, the influence of geometric parameters on turbine performance is not well-understood, in part because there are neither generalized analytical formulations nor inexpensive, accurate numerical models that describe their fluid dynamics. Here,  we experimentally investigate the effect of aspect ratio – the ratio of the blade span to rotor diameter – on the performance of a straight-bladed cross-flow turbine in a water channel. To isolate the effect of aspect ratio, all other non-dimensional parameters are held constant, including the relative confinement, Froude number, and Reynolds number. The coefficient of performance is found to be invariant for the range of aspect ratios tested (0.95 – 1.63), which we ascribe to minimal blade-support interactions for this turbine design. Finally, a subset of experiments is repeated without controlling for the Froude number and the coefficient of performance is found to increase, a consequence of Froude number variation that could mistakenly be ascribed to aspect ratio. This highlights the importance of rigorous experimental design when exploring the effect of geometric parameters on cross-flow turbine performance.
\end{abstract}

\maketitle

\section{Introduction}
Cross-flow turbines convert the kinetic power in wind or water currents to useful mechanical power \citep{Khan2009, Dabiri2011}. While designs with fixed blade pitch have only a single degree of rotational freedom, static geometric parameters have important implications for turbine performance. These parameters include the dimensions of the rotor (e.g., aspect ratio \citep{Brusca2014, Li2017}, helix angle \citep{Shiono2002}, and number of blades \citep{Castelli2012, Li2015}), properties of the blades (e.g., foil profile \citep{SimaoFerreira2013, Bedon2016}, preset-pitch angle \citep{Fiedler2009, Strom2015}, and chord-to-radius ratio \citep{Migliore1980, Balduzzi2015}), and the type of support members used to attach the blades to the central shaft (e.g., struts or end-plates \citep{Strom2018, Villeneuve2019}). 

The breadth of this parameter space suggests optimization by reduced-order models, but this is complicated by the unsteady nature of cross-flow turbine fluid dynamics. The blade's angle of attack varies throughout the turbine’s rotation, potentially inducing dynamic stall \citep{Tsai2016, Buchner2015}, and the blades pass disturbed flow during their downstream sweep, potentially interacting with coherent structures. Such dynamics are difficult to generalize to the full range of rotor geometries. Unlike blade element momentum (BEM) models for axial-flow turbines \citep{Burton2011}, analytical models \citep{Paraschivoiu2002, Islam2008, Mohammed2019} are often only valid for a specific configuration. Similarly, the challenges of simulating unsteady boundary layer dynamics \citep{Tsai2016, Buchner2015}, combined with a lack of exploitable circumferential symmetry and a necessity for high spatial and temporal resolution \citep{Balduzzi2016}, make accurate two-dimensional simulations computationally expensive. Three-dimensional simulations, which are even more expensive, are required to fully incorporate the effects of span-wise flow, mixing, and parasitic torque from blade supports \citep{Ferreira2014,Bachant2016b,Bianchini2017}. 

These limitations create a natural gravity towards experimental exploration of geometric optimization. Here, computational shortcomings are exchanged for limits on turbine size and flow conditions imposed by the test facility (e.g., maximum Reynolds number \citep{Miller2018}). Several experimental studies have investigated the effects of a variety of geometric parameters on cross-flow turbine performance \cite{Shiono2000, Shiono2002, Fiedler2009, Li2015, Strom2015, Bachant2016, Strom2018, Strom2019}. However, experiments must be structured such that the geometric parameter of interest is fully isolated (i.e., all other non-dimensional parameters that affect turbine performance must be held constant). This can be difficult to achieve in practice, and, if neglected, erroneous conclusions may be drawn about the effect of a geometric parameter.

For example, consider the turbine aspect ratio, a geometric parameter defined as
\begin{equation}
    AR = \frac{H}{D},
\end{equation}

\noindent where $H$  is the blade span, and $D$ is the rotor diameter. The aspect ratio has several practical implications for turbine implementation. First, for a cross-flow turbine in a cantilevered orientation, as aspect ratio increases while the diameter is held constant, the moment on the drive train bearings and supporting structure increases, with attendant cost and design complexity. Second, as aspect ratio increases, intermediate support structures may be required between the blades and drive shaft to limit mechanical stress. These intermediate structures can negatively affect power generation \citep{Strom2018}. 

Previous investigations of the effect of aspect ratio on cross-flow turbine performance have focused on simulation, although some experiments have been conducted. \citet{Brusca2014} utilized an analytical model (multiple stream tube) to characterize the effect of aspect ratios from 0 to 3 on the performance of a straight-bladed vertical-axis wind turbine. The authors concluded that a decrease in aspect ratio increased efficiency. This was attributed to an increase in the optimal Reynolds number for lower aspect ratios, which would be expected to increase turbine performance while in a Reynolds-dependent regime \citep{Miller2018}. However, because aspect ratio was varied by changing the turbine diameter, this simultaneously varied aspect ratio along with the chord-to-radius ratio $(c/R)$ and solidity $(Nc / \pi D)$, both of which can significantly affect turbine performance \citep{Strom2019, Shiono2000}. Specifically, as chord-to-radius ratio and solidity decrease, the optimal tip-speed ratio increases, thus increasing the local blade Reynolds number, and overall turbine efficiency for Reynolds-dependent regimes. In addition, as chord-to-radius ratio increases, an empirical correction for flow curvature is required \citep{Migliore1980, Balduzzi2015}. Consequently, it would be inappropriate to ascribe the change in performance solely to the aspect ratio.

\citet{Li2017} investigated the effect of aspect ratio on a straight-bladed cross-flow wind turbine using a three-dimensional simulation (panel method). Turbine aspect ratios from 0.4 to 1.2 were investigated, and both the chord-to-radius ratio and solidity were held constant for all cases. The simulation predicted an increase in efficiency with increasing aspect ratio. The authors attributed this to a decrease in the detrimental effects of blade tip vortices. In other words, as the turbine blade span increased, the tip losses became relatively less significant.

\citet{Hyun2012} experimentally examined the performance of three-bladed and four-bladed cross-flow turbines with aspect ratios of 1.5 and 2 in a circulating water channel with a free surface. As for Li \textit{et al.}, these experiments held the chord-to-radius ratio and solidity constant. While the reported turbine efficiency increased with increasing aspect ratio, this parameter was not fully isolated by the experimental design. The turbines were tested in a water channel with a constant depth, such that the confinement varied with blade span: for $AR = 1.5$, "blockage" (i.e., the ratio of turbine projected area to the channel cross-section) was 30\%, while for $AR = 2$, blockage was 40\%. The authors attributed the improved performance of the higher aspect ratio turbine to the turbine blades piercing the free surface, but the higher confinement would increase efficiency \citep{Garrett2007, Consul2013, Houlsby2017, Ross2020} and ventilation from piercing the surface (i.e., air entrained into the rotor) would reduce lift \citep{Young2017}. Consequently, the efficiency increase attributed to aspect ratio is inextricably convolved with confinement and ventilation.

Consequently, to our knowledge, there has been no experimental investigation that fully isolates the effect of aspect ratio on cross-flow turbine performance. Further, because of convolved non-dimensional variables in both simulation and experimental studies, prior work has drawn conflicting conclusions about the effects of aspect ratio. Here, we evaluate the performance of a straight-bladed cross-flow turbine with blade-end struts over a range of aspect ratios through an experimental method that holds all other relevant non-dimensional parameters constant. In doing so, we provide insight into the effect of aspect ratio on cross-flow turbine performance and the experimental design required to fully isolate the influence of geometric parameters.

\section{Methods}

\subsection{Experimental Set-Up}

\begin{figure} 
    \centering
    \includegraphics[width = 0.9\textwidth]{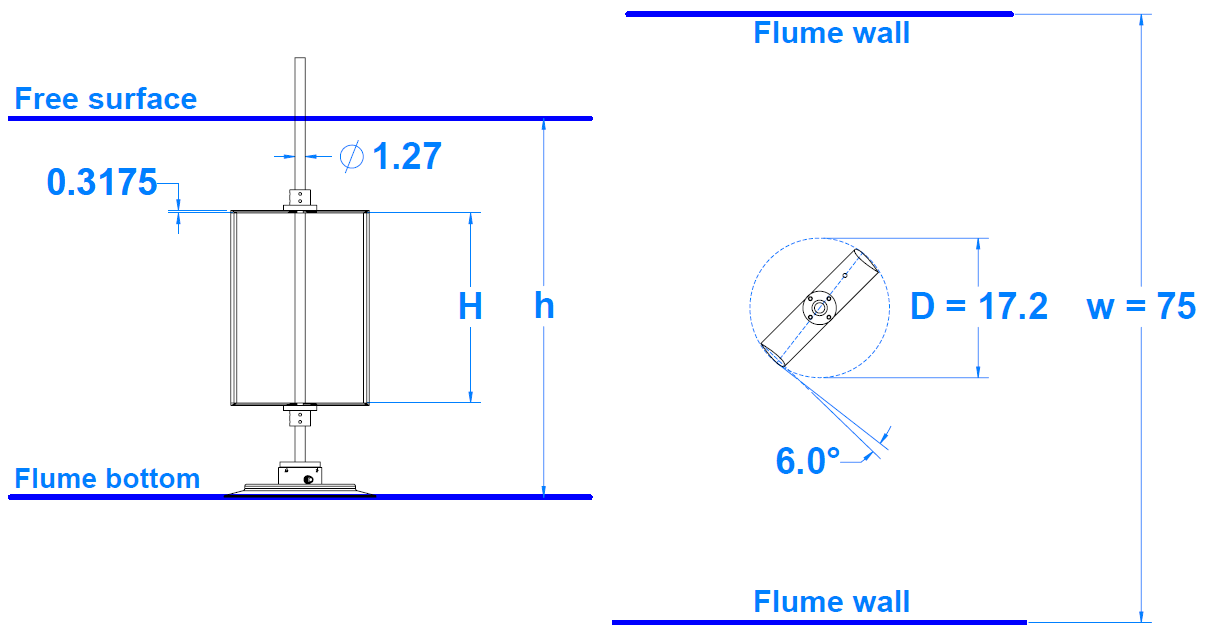}
    \caption{Schematic of the cross-flow turbine used in this study. Aspect ratio was varied by changing the blade span, $H$, while holding the diameter, $D$, constant. The flume width and dynamic depth are given by $w$ and $h$, respectively. Units are centimeters unless otherwise indicated.}
    \label{fig:turbDrawing}
\end{figure}

\begin{figure} 
    \centering
    \begin{minipage}[t]{0.35\textwidth}
        \centering
        \includegraphics[width = 0.9\textwidth]{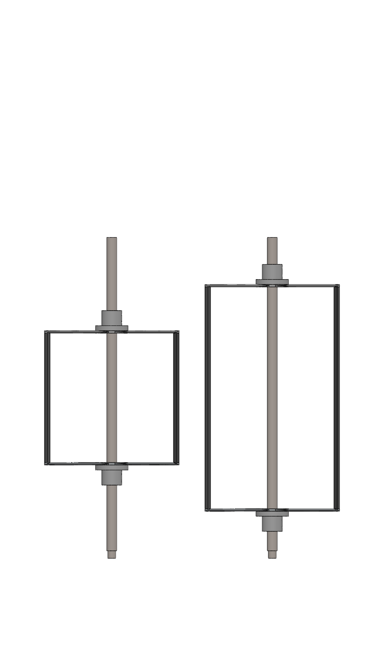}
        \caption{Comparison of the lowest aspect ratio turbine (left, $AR = 0.95$) to the highest (right, $AR = 1.63$).}
        \label{Fig:ARComparison}
    \end{minipage} \hfill
    \begin{minipage}[t]{0.55\textwidth}
        \centering
        \begin{overpic}[width = 0.85\textwidth]{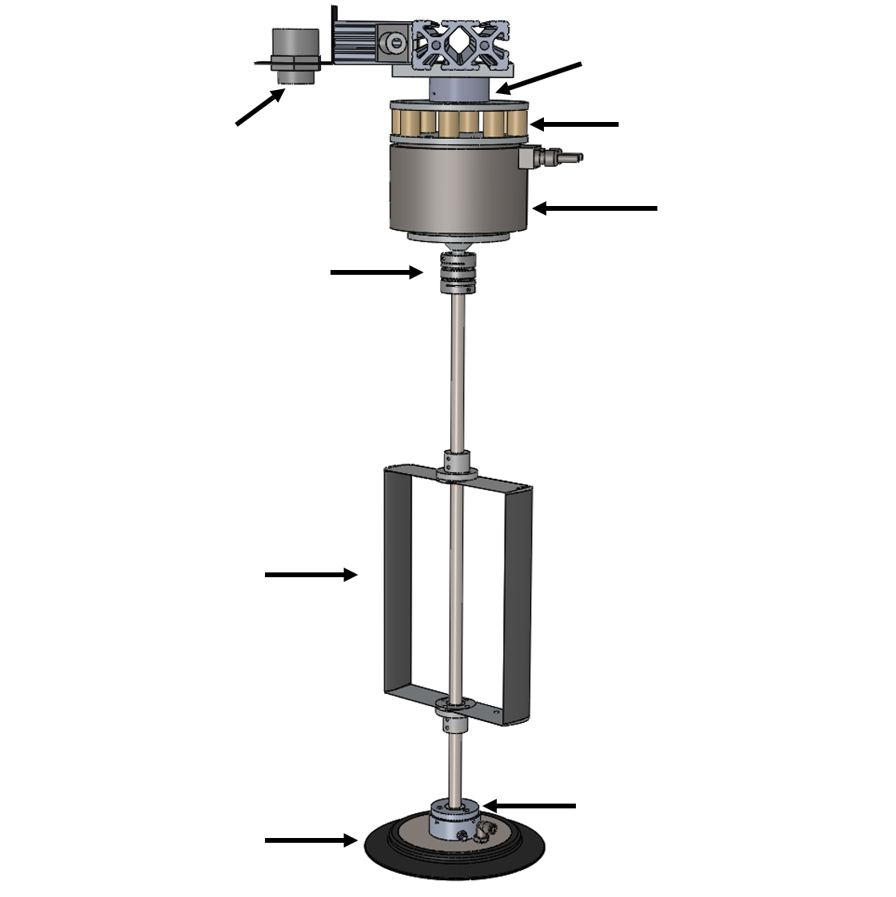}
                \put(-1, 5.4){\small Suction plate}
                \put(65, 9){\small Mini40 load cell}
                \put(11.5, 34.75){\small Turbine}
                \put(5.25, 69){\small Shaft coupling}
                \put(73, 74){\small\setlength\extrarowheight{-20pt}\begin{tabular}{c} Servomotor,\\encoder \end{tabular}}
                \put(70, 85){\small Thermal isolation}
                \put(-1, 84){\small\setlength\extrarowheight{-20pt}\begin{tabular}{c} Free surface\\transducer \end{tabular}}
                \put(65.5, 92.5){\small Mini45 load cell}
        \end{overpic}
        \caption{A rendering of the turbine experimental set-up. The upper load cell is connected to a rigid crossbeam that spans the flume width.}
        \label{fig:expSetUp}
    \end{minipage}
\end{figure}

The experimental turbine used in this study was a cross-flow turbine with two straight blades, as illustrated in \Cref{fig:turbDrawing}. To vary aspect ratio, multiple turbine blades, each with a different blade span, $H$,  were manufactured, while chord length (4.05 cm), blade profile (NACA 0018), turbine diameter (17.2 cm), and preset pitch angle ($6^{\circ}$) were held constant. These parameters correspond to aspect ratios from 0.95 to 1.63 and a constant chord-to-radius ratio of 0.24. For comparison, the turbines with the smallest and largest aspect ratios are shown in \Cref{Fig:ARComparison}. The blades were attached to the central shaft at their ends using thin, hydrodynamic struts (NACA 0008 profile, c\textsubscript{strut} = 4.05 cm) to minimize support structure losses and blade-strut interactions \citep{Strom2018}. Blades and struts were machined from 6061 aluminum, sanded, coated with a thin layer of grey Dupli-Color filler primer, and painted black with Krylon ultra-flat black spray paint.

Experiments were conducted in the Alice C. Tyler recirculating water flume at the University of Washington. The flume is 75 cm wide and has closed-loop temperature control for heating or cooling. The experimental setup, shown in \Cref{fig:expSetUp}, is as described in \citet{Polagye2019} with the exception of the bottom load cell, acoustic Doppler velocimeter, and free surface transducer. Salient details are repeated here.

The top end of the turbine shaft is attached via a shaft coupling to a servomotor (Yaskawa SGMCS-05BC341), which regulates the turbine rotational rate to a constant, prescribed value. The angular position of the turbine is measured by the servomotor encoder, and angular velocity is calculated by differentiation with respect to time. A six-axis load cell (ATI Mini45) attached to the top of the servomotor measures moments and forces on the turbine in the horizontal and vertical directions. The bottom end of the turbine shaft rests in a bearing on a second six-axis load cell (ATI Mini40). The bearing and load cell are affixed to the bottom of the flume using a suction plate. Measurements from the encoder and load cells were logged by a National Instruments PCIe-6351 card through a custom Simulink model in MATLAB at a sample rate of 1000 Hz.

The inflow velocity was measured using an acoustic Doppler velocimeter (Nortek Vectrino profiler). A single cell, centered vertically and horizontally in the water column, was sampled at 64 Hz at a location 5 turbine diameters upstream from the center shaft. Velocity measurements were de-spiked using the method of \citet{Goring2002}. The dynamic depth of the water channel was measured approximately $0.50D$ upstream and $0.75D$ laterally offset from the center of the turbine using a free-surface transducer (Omega LVU30 series) sampling at 10 Hz. The water temperature was measured using a temperature probe (Omega T-type) with a resolution of $\pm 0.1 \, ^\circ$C and logged manually every minute during each experiment. 

\subsection{Non-dimensional Parameters}
Because the confinement \citep{Garrett2007, Consul2013, Houlsby2017, Ross2020}, Froude number \citep{Houlsby2017, Consul2013, Gauvin-Tremblay2020}, and Reynolds number \citep{Bachant2016a, Miller2018}, can all affect turbine performance, we designed the experiments to hold these constant. The confinement, or blockage ratio, experienced by the turbine is defined as the ratio of the turbine’s projected area to the cross-sectional area of the channel
\begin{equation} 
    \beta = \frac{A_{turbine}}{A_{channel}} = \frac{HD}{hw},
\end{equation}

\noindent where $h$ is the dynamic depth of the water in the flume, and $w$ is the width of the flume. During testing, each turbine was centered vertically and horizontally in the water column, and the water depth was adjusted to maintain a constant blockage ratio for all turbine geometries. Therefore, as aspect ratio was increased by increasing turbine height $(H)$, the dynamic depth $(h)$ was also increased. However this change in water depth decreased the Froude number, which is given by
\begin{equation}
    Fr = \frac{U_{0}}{\sqrt{gh}},
\end{equation}

\noindent where $U_{0}$ is the inflow velocity, and $g$ is the acceleration due to gravity. One interpretation of $Fr$ is the ratio of the flow’s kinetic energy to its potential energy. To maintain $Fr$ while increasing the water depth, $U_{0}$ was increased. However, increasing $U_{0}$ increased the Reynolds number, defined with respect to the blade chord length as
\begin{equation}
    Re = \frac{U_{0}c}{\nu},
\end{equation}

\noindent where $c$ is the turbine blade chord length, and $\nu$ is the kinematic viscosity of the working fluid. $Re$ represents the ratio of inertial forces to viscous forces in a fluid flow. To hold $Re$ constant, the kinematic viscosity was increased by decreasing the water temperature ($T$). In this manner, we were able to hold all three parameters constant while varying only the aspect ratio. While $Re$ can be defined in terms of other length or velocity scales, this definition has the advantage of simplicity. For example, incorporating turbine rotation into the velocity scale yields variations in $Re$ with rotation rate and still neglects induction. As our focus is on rotor hydrodynamics, the chord length was chosen as the length scale, but is proportional to the diameter.

Experiments were conducted at a blockage ratio of 11.5\% and at two $Re$-$Fr$ pairs: one "high pair" $(Re = 4.27 \times 10^{4},\, Fr = 0.427)$ and one "low pair" $(Re = 2.03 \times 10^{4},\, Fr = 0.279)$. This was done to confirm that trends observed at one combination of operating parameters were consistent with another. The selected values of $Re$ and $Fr$ maximize the range of $Re$ achievable in the Tyler flume given the facility constraints on $h$, $U_{0}$, and $T$, and the value for $\beta$ was chosen to be in the range of previous experiments \cite{Strom2019}. Additionally, since controlling the Froude number in cross-flow turbine performance experiments is not a universal practice, the high $Re$-$Fr$ tests were repeated while holding only confinement and $Re$ constant to explore the significance of $Fr$ on results. The values of $h$, $U_{0}$, and $T$ required to maintain $\beta$, $Fr$ (when held constant), and $Re$ for each aspect ratio are given in \Cref{table:expParams}. The turbulence intensity ($TI$) for each case is also reported. As turbine performance has been shown to be sensitive to $TI$ primarily at higher levels \citep{Maganga2010, Blackmore2016}, we do not expect $TI$ to significantly affect our results.

\begin{table}
    \centering
    \caption{Turbine geometries, flume conditions, and resulting non-dimensional parameters for the three sets of experiments conducted.}
    \label{table:expParams}
    \setlength{\extrarowheight}{-10pt} 
    \setlength{\tabcolsep}{6pt}        
    \resizebox{\textwidth}{!}{      
        \begin{tabular}{@{}ccccccccc@{}}
            \toprule
            \textbf{Experiment} & \textbf{Turbine} & \multicolumn{4}{c}{\textbf{Nominal Flume Parameters}} & \multicolumn{3}{c}{\textbf{\begin{tabular}[c]{@{}c@{}}Nominal Non-Dimensional\\ Flow Characteristics\end{tabular}}} \\
            \cmidrule(lr){1-1} \cmidrule(lr){2-2} \cmidrule(lr){3-6} \cmidrule(lr){7-9}
             & $AR$ & $h$ (cm) & $U_{0}$ (m/s) & $T$ $(^{\circ} $C) & $TI$\footnotemark[1] (\%) &
             $B \, (\%)$ & $Fr$ & $Re$ \\ 
             \cmidrule(lr){2-2} \cmidrule(lr){3-3} \cmidrule(lr){4-4} \cmidrule(lr){5-5} \cmidrule(lr){6-6} \cmidrule(lr){7-7} \cmidrule(lr){8-8} \cmidrule(lr){9-9}
            \multirow{6}{*}{High $Re$-$Fr$} & 0.95 & 32.7 & 0.764 & 35.0 & 1.7 & \multirow{6}{*}{11.5} & \multirow{6}{*}{0.427} & \multirow{6}{*}{$4.27\! \times\!10^{4}$} \\
             & 1.09 & 37.3 & 0.816 & 31.7 & 1.6 &  &  &  \\
             & 1.22 & 42.0 & 0.866 & 28.9 & 1.6 &  &  &  \\
             & 1.36 & 46.7 & 0.913 & 26.4 & 1.6 &  &  &  \\
             & 1.50 & 51.3 & 0.957 & 24.3 & 1.6 &  &  &  \\
             & 1.63 & 56.0 & 1.000 & 22.4 & 3.3 &  &  &  \\ \midrule
            \multirow{6}{*}{Low $Re$-$Fr$} & 0.95 & 32.7 & 0.500 & 20.2 & 2.2 & \multirow{6}{*}{11.5} & \multirow{6}{*}{0.279} & \multirow{6}{*}{$2.03\! \times\!10^{4}$} \\
             & 1.09 & 37.3 & 0.535 & 17.5 & 2.2 &  &  &  \\
             & 1.22 & 42.0 & 0.567 & 15.2 & 1.9 &  &  &  \\
             & 1.36 & 46.7 & 0.598 & 13.3 & 1.7 &  &  &  \\
             & 1.50 & 51.3 & 0.627 & 11.5 & 1.6 &  &  &  \\
             & 1.63 & 56.0 & 0.655 & 10.0 & 1.4 &  &  &  \\ \midrule
            \multirow{6}{*}{\begin{tabular}[c]{@{}c@{}}High $Re$,\\ variable $Fr$\end{tabular}} & 0.95 & 32.7 & \multirow{6}{*}{1.000} & \multirow{6}{*}{22.4} & 1.5 & \multirow{6}{*}{11.5} & 0.559 & \multirow{6}{*}{$4.27 \! \times \! 10^{4}$} \\
             & 1.09 & 37.3 &  &  & 1.5 &  & 0.523 &  \\
             & 1.22 & 42.0 &  &  & 1.6 &  & 0.493 &  \\
             & 1.36 & 46.7 &  &  & 1.7 &  & 0.467 &  \\
             & 1.50 & 51.3 &  &  & 1.8 &  & 0.446 &  \\
             & 1.63 & 56.0 &  &  & 3.3 &  & 0.427 &  \\ 
             \bottomrule
        \end{tabular}
    }
    \footnotetext[1]{Defined as $\frac{\sqrt{2k/3}}{U_{0}}$, where $k$ is the turbulence kinetic energy of the inflow.}
\end{table}

\subsection{Performance Metrics}
The efficiency of the turbine, also known as the coefficient of performance or power coefficient, is the ratio of power produced by turbine to the kinetic power in the fluid that passes through the rotor swept area. The time-average coefficient of performance is given by
\begin{equation}
    C_{P} \equiv \left\langle C_{P} \right\rangle = 
    \left\langle \frac{\tau(t) \omega(t)} 
    {\frac{1}{2} \rho A U_{0}(t)^{3}} \right\rangle,
\end{equation}

\noindent where $\tau(t)$ is the instantaneous hydrodynamic torque on the turbine, $\omega(t)$ is the turbine’s instantaneous angular velocity, $U_{0}(t)$ is the instantaneous freestream velocity, and $\rho$ is the density of the working fluid. A Butterworth filter was applied to the torque measurements to eliminate motor noise from the data. Because the ADV velocity measurement was not synchronized with other data acquisition, an advection time correction was applied to the inflow velocity \citep{Polagye2019}.

$C_{P}$ varies with the turbine’s tip-speed ratio, which is the ratio of the turbine blade’s tangential velocity to the freestream velocity, given as
\begin{equation}
    \lambda = \frac{\omega R}{U_{0}},
\end{equation}

\noindent where $R$ is the radius of the turbine rotor. Turbines with each aspect ratio were tested from $\lambda = 0.8$ to $\lambda = 3.2$ with a step size of 0.1. Data was collected at each tip-speed ratio for 60 seconds, and the measurement time series were trimmed to an integer number of turbine rotations.

Additionally, to isolate the performance of the turbine blades from the rest of the rotor (i.e., the struts and central shaft), the parasitic support structure torque was measured by repeating each experimental condition with a bladeless turbine. The performance of the blades was then estimated as:
\begin{equation}
    C_{P, blade} \cong C_{P, turbine} - C_{P, supports}.
\end{equation}

\noindent This strategy, first proposed by \citet{Bachant2016}, assumes that secondary interactions between the blades and support structures are limited, which was borne out by \citet{Strom2018} for a range of support structures, including the struts used here.

\section{Results}

\begin{figure} 
    \centering
    \includegraphics[scale = 1]{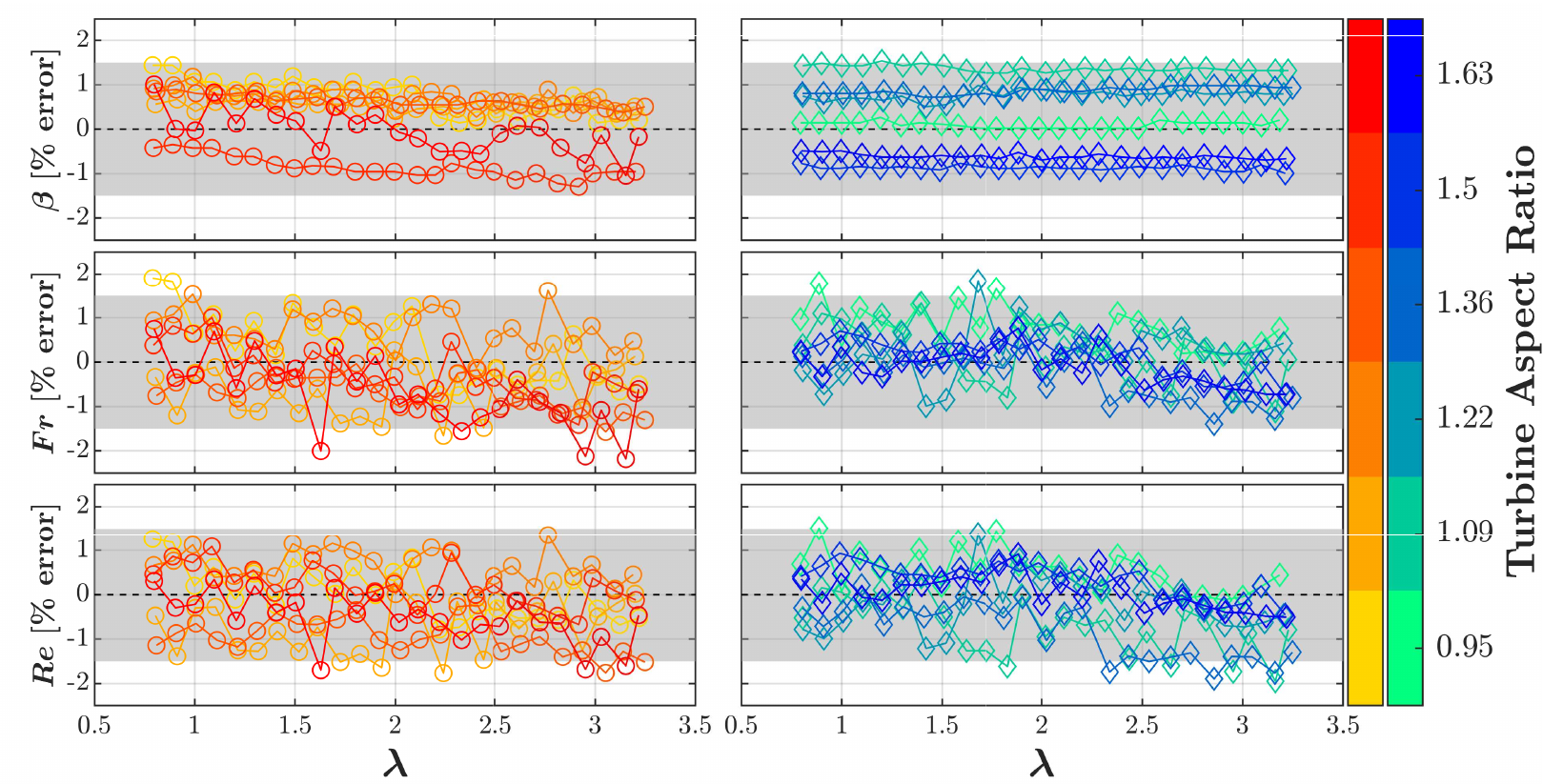}
    \caption{Percent error between target and measured time-average values for the blockage ratio, Froude number, and Reynolds number shown for the high $Re$-$Fr$ experiments (left) and the low $Re$-$Fr$ experiments (right). The grey shading denotes $\pm1.5\%$ deviation from the target value.}
    \label{fig:flowParams}
\end{figure}

\subsection{Parameter Control}
Across experiments, $\beta$, $Fr$, and $Re$ were maintained within approximately $\pm1.5\%$ of target values, as shown in \Cref{fig:flowParams}. The relatively small variation in these parameters for a given aspect ratio can be attributed to long-period ($>$ 10 s) oscillations corresponding to resonance between the pumps and flume structure. The slight decrease in $\beta$, $Fr$, and $Re$ with increasing $\lambda$ is due to greater rotor thrust at higher angular velocities, which elevates the free-surface ahead of the rotor and reduces the freestream velocity.

\subsection{Turbine Efficiency}
\begin{figure} 
    \centering
    \includegraphics[scale = 1]{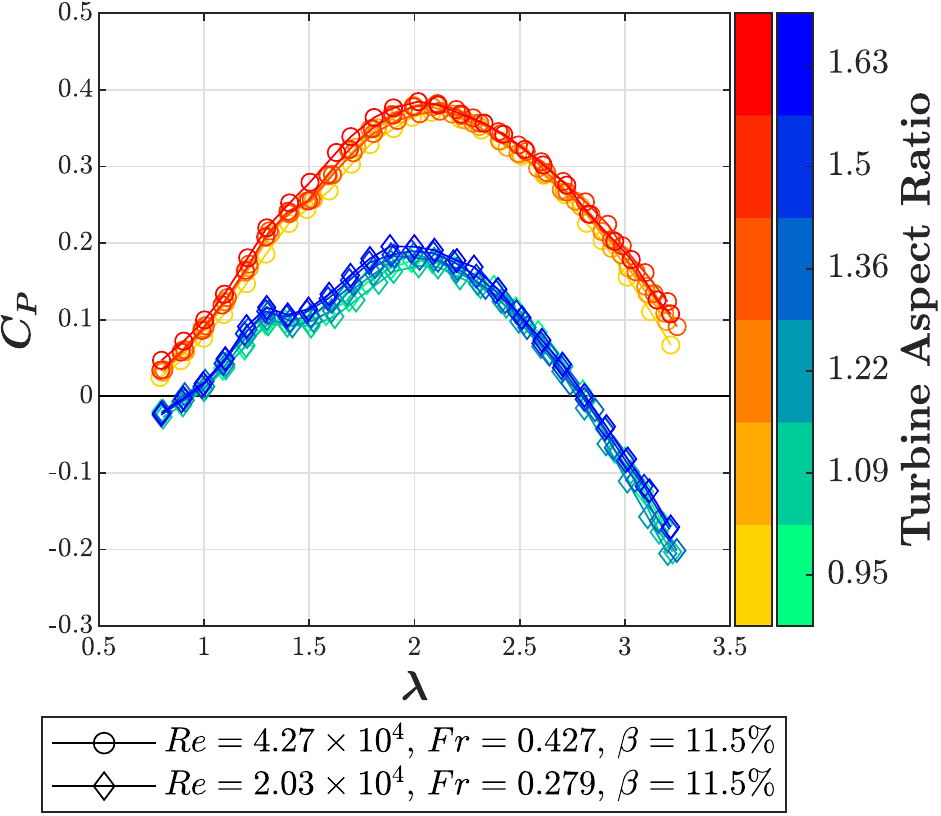}
    \caption{Time-average turbine efficiency versus tip-speed ratio across a range of turbine aspect ratios for two $Re$-$Fr$ combinations.}
    \label{fig:cp3ParamFullTurb}
\end{figure}

As expected, turbine efficiency $(C_{P})$ depends on the $Re$-$Fr$ combination, but is insensitive to aspect ratio (\Cref{fig:cp3ParamFullTurb}). In other words, if all other relevant non-dimensional parameters are held constant, the range of aspect ratio has no effect on this turbine design. The increase in turbine performance with increasing $Re$ at each aspect ratio is expected, as the turbines were operated in a $Re$-dependent regime\citep{Bachant2016a, Miller2018}.

\begin{figure} 
    \centering
    \includegraphics[scale = 1]{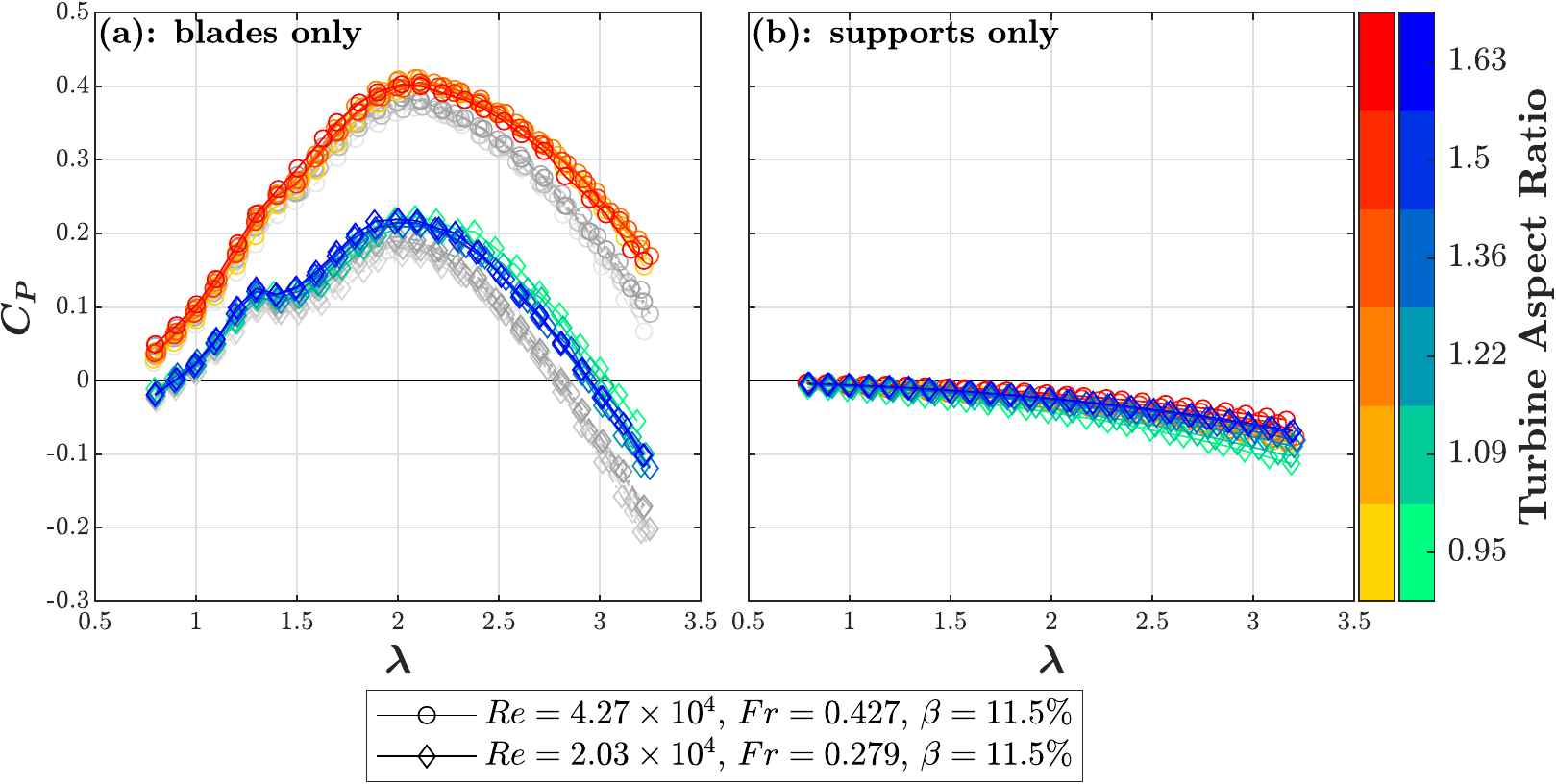}
    \caption{Support structure losses \textbf{(b)} are subtracted from the turbine efficiency (\Cref{{fig:cp3ParamFullTurb}}) to estimate \textbf{(a)} blade-only $C_P$. The grey curves in \textbf{(a)} show the full turbine performance from \Cref{fig:cp3ParamFullTurb}.}
    \label{fig:cp3Subtraction}
\end{figure}

The scatter in peak $C_{P}$ (defined as $\Delta C_{P}$) for the $Re$-$Fr$ combinations was 0.01 and 0.03 for the higher and lower combination, respectively. As shown in \Cref{fig:cp3Subtraction}a, this is further reduced when the performance of the blade is isolated by subtracting the losses from the support structures (\Cref{fig:cp3Subtraction}b). We ascribe the residual scatter in the lower $Re$-$Fr$ combination to low signal-to-noise ratio for torque.

\subsection{Froude Number Importance}
\begin{figure} 
    \centering
    \includegraphics[scale = 1]{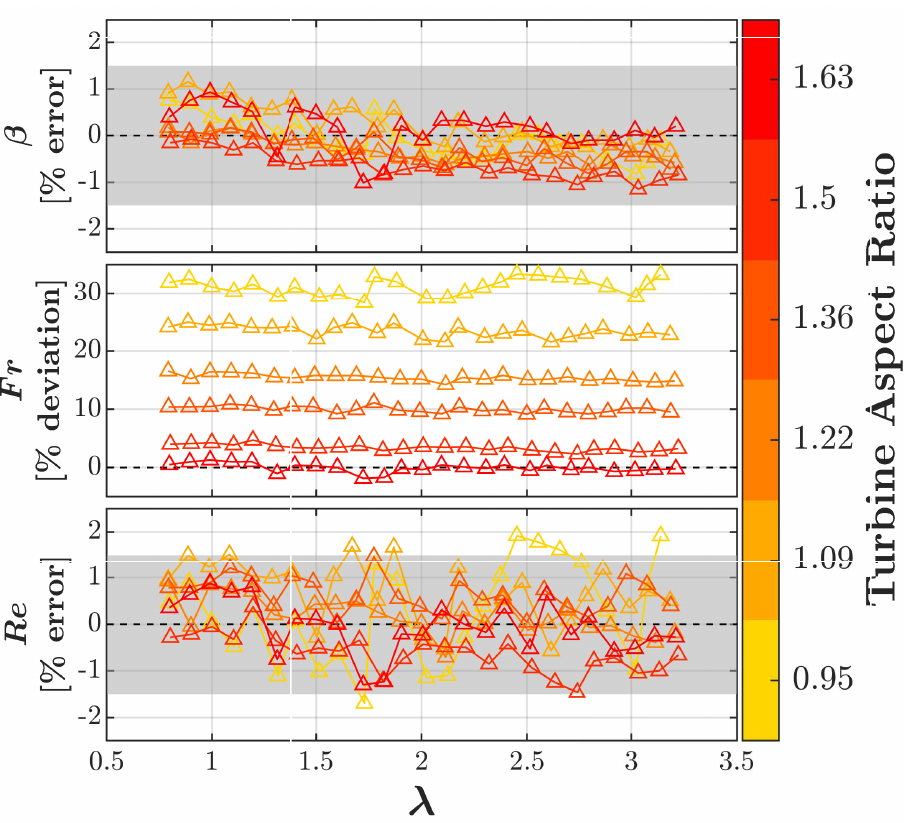}
    \caption{Percent deviation between target and measured time-average values of the blockage ratio, Froude number, and Reynolds number for the high $Re$ case when $Fr$ is allowed to vary. For $\beta$ and $Re$, the grey shading denotes $\pm 1.5\%$ deviation from the target values.}
    \label{fig:flowParamsFr}
\end{figure}

\Cref{fig:flowParamsFr} shows the percent deviation from target values of $\beta$ and $Re$ for high $Re$ experiments repeated at varying $Fr$. As a consequence of varying water depth, the Froude number increases by more than 30\% from the largest to smallest aspect ratio turbine.

\begin{figure} 
    \centering
    \includegraphics[scale = 1]{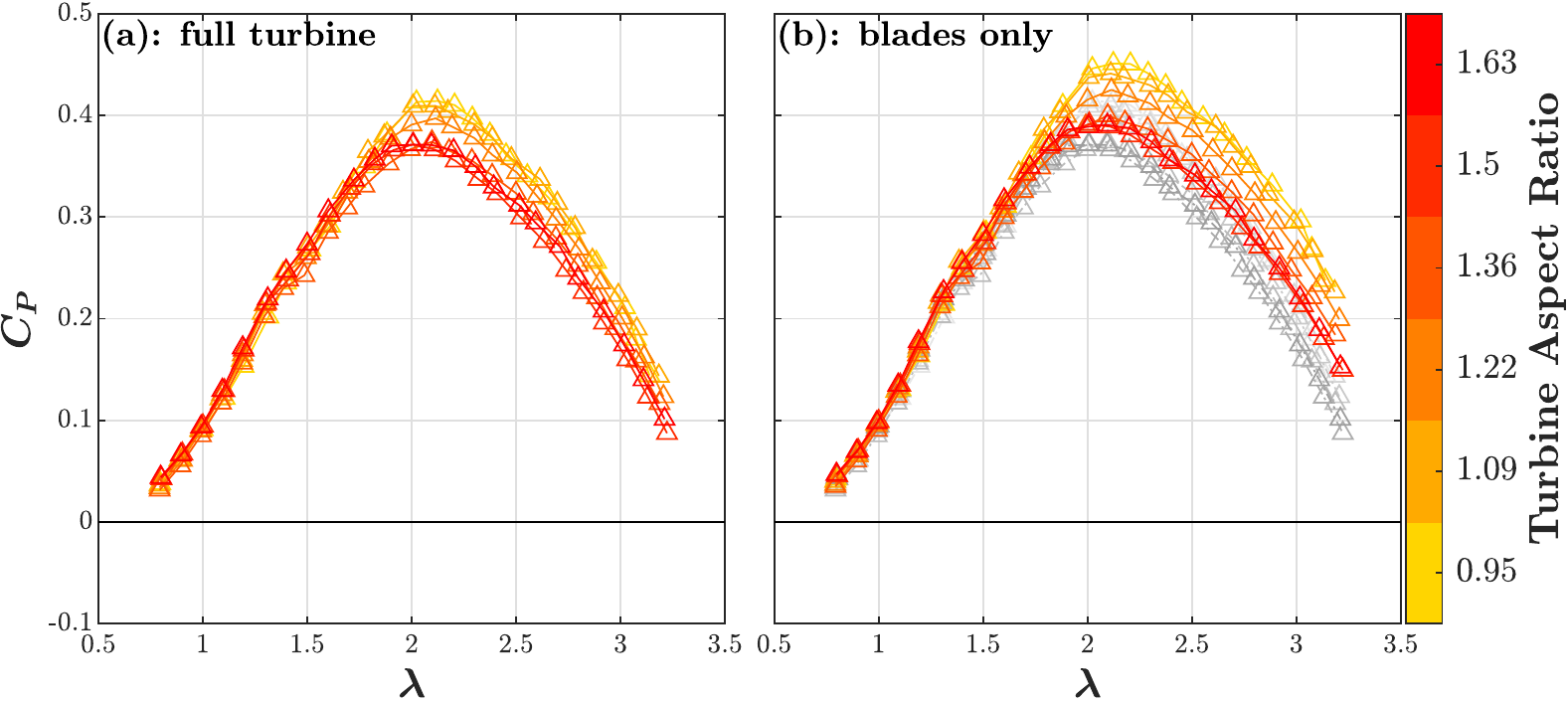}
    \caption{Time-average $C_{P}$ versus tip-speed ratio across aspect ratio for \textbf{(a)} the full turbine and \textbf{(b)} the turbine blades only for the high $Re$ and variable $Fr$. The grey curves in \textbf{(b)} show the full turbine performance from \textbf{(a)}.}
    \label{fig:cp2Subtraction}
\end{figure}

\Cref{fig:cp2Subtraction}a shows how $C_{P}$ increases substantially with decreasing aspect ratio and increasing $Fr$. This trend is amplified for the blade-level performance coefficient in \Cref{fig:cp2Subtraction}b. As we have already demonstrated that performance is independent of aspect ratio when $Fr$ is held constant, the observed change is solely a consequence of increasing $Fr$. The increased efficiency with $Fr$ is consistent with simulations of cross-flow turbines by \citet{Consul2013} and \citet{Gauvin-Tremblay2020}.

\begin{figure} 
    \centering
    \includegraphics[width = 0.7\textwidth]{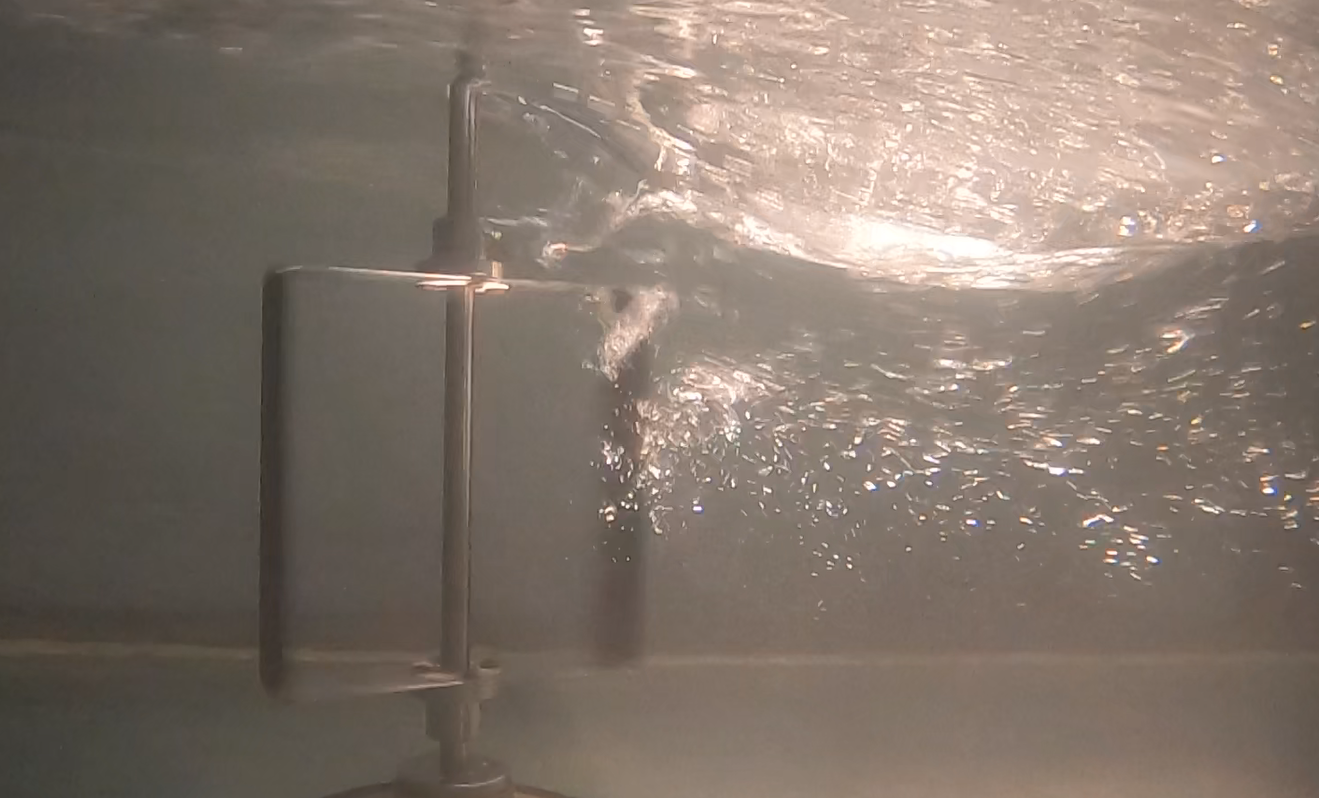}
    \caption{The most severe ventilation observed on the downstream sweep of the blade ($AR = 0.95$ for the variable $Fr$ case).}
    \label{fig:ventilationPic}
\end{figure}

As a consequence of holding the blockage ratio constant for these experiments, when the aspect ratio was decreased, the dimensional distance between the turbine and the free-surface was also decreased. For experiments with variable $Fr$, we observed increasing ventilation of the downstream turbine blade (see \Cref{fig:ventilationPic}). Although the torque contributed by the downstream blade is small, ventilation would be expected to increase the form drag on the blade \citep{Young2017} and, consequently, reduce the net rotor torque. Therefore, the performance increase with $Fr$ may have been even larger if not for ventilation losses.

\section{Discussion}
\subsection{Effect of Aspect Ratio on Performance}
We find that, when other non-dimensional parameters are held constant, aspect ratio has no effect on $C_P$ in the range of $Re$, $Fr$, $\beta$, and $AR$ tested. This outcome is consistent with the underlying fluid mechanics. The only mechanism for aspect ratio to affect turbine performance is through the relative extent of the blade span affected by tip losses or support effects. In our experiments, strut and tip losses were relatively minor compared to power production by the blades (\Cref{fig:cp3Subtraction}). With further reductions in $AR$, we would eventually expect to see $C_{P}$ decline as those losses become appreciable. 

Additionally, with a different blade support architecture, we could anticipate more significant effects at higher aspect ratio. For example, mid-span struts have significant, undesirable blade-strut interactions \citep{Strom2018}, shown by \citet{Villeneuve2019b} to be a consequence of inopportune vortex shedding at the blade-strut interface. \citet{Li2017} employed mid-span struts and, consequently, found performance to increase with aspect ratio, since as the blade span increased, relatively less of the blade was impacted by this phenomenon. Our conclusion that changes in turbine performance with aspect ratio are primarily a consequence of strut interactions is, therefore, consistent with Li \textit{et al.}, who were also rigorous in holding other non-dimensional parameters constant in their simulations. Consequently, we would expect the efficiency of turbines with blade-end supports that produce higher parasitic torque, such as the end-plates and thicker struts \citep{Strom2018}, to increase with aspect ratio.
\subsection{Assessment of Experimental Approach}

The support structure interactions responsible for changes in performance with aspect ratio are inherently three-dimensional. Therefore, an analytical or numerical study exploring this parameter space must either be three-dimensional or include empirical models for tip and support losses to extend a two-dimensional study. Because of this, the experimental approach taken here is a cost-effective exploratory tool. However, in experiments, unlike in simulation, it can be challenging to hold all other relevant non-dimensional parameters constant. By controlling $\beta$, $Fr$, and $Re$, we observed non-dimensional performance similarity (i.e., consistent $C_P$-$\lambda$ dependence) across the range of aspect ratios tested. While this approach limited the range of $AR$ that could be explored, failing to do so would have led to erroneous conclusions about the effects of $AR$. As demonstrated, if the Froude number was allowed to vary, we could have incorrectly attributed an increase in turbine efficiency to a decrease in aspect ratio. This would have been a baffling result defying physical intuition. More broadly, this demonstrates that the Froude number is an important consideration for cross-flow turbine experiments conducted in an open channel. We note that there remains some ambiguity in the appropriate definition of $Fr$, as $Fr$ defined in terms of the separation between the top of the rotor and free surface also remained constant during our experiments.

Finally, rotor ventilation, particularly of the downstream sweep, was observed in some experiments, but not explicitly accounted for in any of the non-dimensional parameters we held constant. As ventilation was most severe for combinations of relatively high flow speeds and low water depth, this is likely to be an important consideration at relatively high Froude numbers. The non-dimensional similarity in turbine performance when $\beta$, $Fr$, and $Re$ were held constant demonstrates that ventilation, even when observable, does not necessarily have a meaningful effect on turbine performance. Further investigation of rotor ventilation on cross-flow turbine performance may be appropriate.

\section{Conclusion}

Optimizing cross-flow turbine geometry could increase their cost-effectiveness. However, to draw firm conclusions about the effects of a specific geometric variable, all other non-dimensional parameters that impact turbine performance must be held constant. Due to a lack of generalized analytical models and the computational expense of three-dimensional numerical simulations, experimental exploration is an attractive approach for exploring geometric variability. Here, as an example, we experimentally assess the effect of rotor aspect ratio on turbine performance while rigorously maintaining the confinement, Froude number, Reynolds number, and other non-dimensional rotor parameters (e.g., chord-to-radius ratio).

For the straight-bladed cross-flow turbine considered here, performance is shown to be independent of aspect ratio from 0.95 to 1.63 when other non-dimensional parameters are held constant. The mechanism for changes in turbine performance with aspect ratio is the balance between losses due to blade supports or blade tip vortices and power production by the blades. Our choice of supports (blade-end struts with a thin, foil cross-section) contributed to the aspect ratio invariance for the turbine. When support structure losses are accounted for, blade-level performance is almost entirely independent of aspect ratio. Aspect ratio is, therefore, most likely to affect turbine performance when support structures incur greater losses, such as with end-plates, or produce undesirable blade-strut interactions, such as with mid-span supports. At even lower aspect ratios than those studied here, we would expect support structures to have more pronounced effects since, as the aspect ratio approaches zero, effects on blade-level performance through blade-strut interactions would no longer be negligible.

This study demonstrates the importance of isolating the effects of non-dimensional parameters from the geometric parameter under investigation. In our experiments, we held confinement, Froude number, and Reynolds number constant across all aspect ratios. However, decreasing aspect ratio while holding only confinement and Reynolds number constant results in an increasing Froude number and an associated increase in turbine efficiency. The magnitude of this effect, a turbine efficiency increase of 12\% for a Froude number increase of 30\%, highlights the importance of controlling for this parameter in experiments.

\section*{Acknowledgements}
This work was supported by the United States Department of Defense Naval Facilities Engineering Command (NAVFAC) under contract N0002418F8702. The authors would like to thank Paul Gibbs for prompting the inception of this study, Eamon McQuaide and Reginald Rocamora for their assistance in manufacturing the turbine blades used in these experiments, and Erik Skeel for machining the struts. A special thanks is extended to Benjamin Strom and Hannah Ross for designing and continually improving the cross-flow turbine experimental set-up that was used in this investigation.

\section*{Data Availability}
Turbine performance data supporting this study, as well as a video of the observed rotor ventilation, can be accessed at ResearchWorks, the University of Washington's digital repository. This material may be accessed at \textcolor{blue}{http://hdl.handle.net/1773/45618}.

\bibliography{AR_Paper_Refs.bib}

\end{document}